\documentclass{article}
\usepackage{url}
\usepackage{fancyheadings}
\usepackage[left=2.5cm,top=2cm,right=2.5cm]{geometry}
\usepackage{amsmath}
\usepackage{amsfonts}
\usepackage{amssymb}
\usepackage{amsthm}
\usepackage{natbib}
\usepackage{graphicx}

\usepackage{setspace}
\singlespace
\begin{document}
\author{Lucy F. Robinson$^1$ and Carey E. Priebe$^2$}

\date{}

\title{Detecting Time-dependent Structure in Network Data via a New Class of Latent Process Models}

\maketitle
1. Department Epidemiology and Biostatistics, Drexel University 2. Department of Applied Mathematics and Statistics, Johns Hopkins University 
\\

\begin{abstract}
We introduce a new class of latent process models for dynamic relational network data with the goal of detecting time-dependent structure. Network data are often observed over time, and static network models for such data may fail to capture relevant dynamic features.  We present a new technique for identifying the emergence or disappearance of distinct subpopulations of vertices. In this formulation, a network is observed over time, with attributed edges appearing at random times.  At unknown time points, subgroups of vertices may exhibit a change in behavior. Such changes may take the form of a change in the overall probability of connection within or between subgroups, or a change in the distribution of edge attributes. A mixture distribution for latent vertex positions is used to detect heterogeneities in connectivity behavior over time and over vertices. The probability of edges with various attributes at a given time is modeled using a latent-space stochastic process associated with each vertex. A random dot product model is used to describe the dependency structure of the graph. As an application we analyze the Enron email corpus. \end{abstract}

\section{Introduction}

Network data are often observed over time, with evolving connections between nodes. Our goal is to detect time-dependent structure in network data, finding change points at which the characteristics of the network are altered. We consider ``streaming" network data,  a system in which connections between $n$ vertices are observed over a time period $(0, T)$.   At random times $(t_1, t_2, \ldots, t_N)$, undirected connections between vertices are created, each of which may be associated with a categorical attribute contained in $\{1, 2, \ldots, K\}$.  In particular, we study connections which have no temporal duration, such as emails. At unknown time points, subgroups of vertices may exhibit a change in behavior. Such  changes may take the form of a change in the overall probability of connection within or between subgroups, a change in the distribution of edge attributes, or both. Examples of emerging subpopulations with distinct behavior may include bursts of ``chatter" activity among a group of individuals, or the creation of new communities of vertices such as parents of members of a little league team at the beginning of a new season.

Real world networks, particularly social networks, generally exhibit structure beyond that which can be explained by simple stochastic models. Common features of network structure include heterogeneity across vertices with respect to  number and types of connections observed, transitivity, and clusters of vertices which are more likely to be connected to one another.  One approach to modeling network structure is to partition the graph, grouping together similar vertices. Similarity can be defined in various ways. Many analyses of network data focus on discovering community structure, in which groups of vertices that have a high degree of connection between them belong to the same community or block \citep{gervannewman, airoldifeinbergxing2008}. Vertices can also be classified according to structural equivalence. A group of structurally equivalent vertices have the same ``role" in the network, exhibiting similar patterns of connectivity with other groups of vertices.  Stochastic block models \citep{wangwong, snijdersnowicki} parameterize block structure by assigning within and between-block probabilities of connection, and weighted membership vectors specifying to which blocks a vertex is likely to belong.  

 Latent space models  characterize vertex behavior by  projecting each vertex onto a low-dimensional space. This approach is fruitful both in constructing analytically tractable models, and in creating a framework for comparison across vertices.  \citet{hoffrafteryhandcock} propose a class of latent space models in which the probability of connection  between two vertices depends on their respective latent positions through a Euclidian distance function. Vertices  with neighboring latent positions are more likely to be connected than those with distant positions. After conditioning on latent vertex positions, the edges are independent of one another. A detailed review of statistical network modeling is available in \citet{goldenbergreview}.

Models which define the probability of an edge for a given pair of vertices as a function of latent positions have essentially two possible formulations. The first is a  general latent domain in which unconstrained functions of positions (such as distances) are transformed to probabilities using a link function;  the second is a constrained latent space which produces valid edge probabilities without transformation via a link function. 

We take the latter approach, defining the latent space as the $K$-dimensional simplex and using the inner product between latent positions to define the probability of connection. This ``random dot product model" for undirected connections is described by \citet{scheinermantucker}. The properties of this model with respect to random graph properties such as clustering, diameter and degree distributions are explored in \citet{youngscheinerman}. In a related model, \citet{hoff2009} describes the probability of directed connections using a multiplicative latent factor model in which each vertex is endowed with unobserved sender and receiver characteristics.  

Graph partitions can be defined using the similarity measure induced by distances in latent space.  \citet{handcockraftery2007} apply a  model-based clustering algorithm  to latent positions in a Euclidean space to infer a latent cluster structure over vertices. Latent positions are assumed to be drawn from a Gaussian mixture, and the probability of connection varies with the pairwise distance between latent positions. The latent mixture model induces clustering in the observed network. 

We are interested in estimating  partitions that identify dissimilarity in connectivity behavior across time as well as across vertices. While most network analysis assumes a static topology, dynamic network models have been proposed in contexts where data on the temporal location of connections is available. Dynamic mixed-membership stochastic blockmodels \citep{xingfusong2010}  are an extension of the mixed-membership stochastic blockmodel \citep{airoldifeinbergxing2008}  in which the vectors identifying block membership (for each node) can change over time, such that each actor may take on various roles in a network at different times. 
 Latent space models have also been generalized to include dynamic positions \citep{ sarkarmoore2008, westfieldhoff2010,  xuzheng2009} .  We are particularly interested in dynamic position models which can be analyzed to detect change points at which there is a shift in network behavior. 

In many applications observed edges  contain additional information on the type of connection between two vertices.  We develop a novel treatment of  the ``attributed edge" case (sometimes called  ``colored" edges)  in which each edge has a categorical attribute contained in a finite set, in addition to our  model for  the standard unattributed edge case. Attributes may be observed directly or, as in the case of the Enron email data, assigned by a classification procedure \citep{priebepark2010}. For data with attributed edges, we use the additional information to construct a more complex description of vertex behavior, in which vertices vary with respect to the types of connections they are associated with as well as probabilities of connection. 

\citet{leepriebe2010} propose a dynamic dot product model for attributed graphs in which each vertex is associated with a latent stochastic process.  Using a mathematically tractable approximate model, they develop a test statistic for detecting changes in attributed multigraphs observed at discrete times. The changes of interest are  shifts in the behavior of an unknown anomalous subgroup of vertices.  We approach a similar problem using a related latent position model, and extend the detection problem to multiple change points and groups of  vertices. \citet{costakulldorff2007} investigate a related detection task in a different context, developing a scan statistic to detect spatio-temporal disease clusters. 

We propose a dynamic latent position model in which the presence and attributes of observable edges depend probabilistically on the latent positions of the associated vertices through the random dot product model.  Through the inferred latent positions, we can identify dissimilarity in behavior across groups of vertices or through time. We introduce a mixture model in which latent positions are drawn from either a homogeneous or heterogeneous distribution with respect to vertices and time.  The goal of the analysis is to detect change points at which there is an identifiable shift between heterogeneity and  homogeneity (in either direction) in the network. We approach the possibility of multiple change points through an iterative partitioning procedure. 

Network data are complex. We aim to construct a simple model which can be used to detect time-varying structure in a variety of datasets.  If more subtle features of the network are of interest, an initial evaluation of time-dependent block structure may be useful in segmenting the data before fitting a more elaborate model. As an application we analyze the Enron email corpus.

\section{Model}
\label{sec:model}
We introduce a generative model for streaming network data in which a population of $n$ vertices are observed continuously in time, with edges appearing at random times, possibly with categorical attributes. The density of edges over time is of interest, as is the density of edges across the population of vertices. The distribution of edges over time is assumed to follow a doubly stochastic Poisson process model. 
The generative model is as follows: 

\begin{enumerate}
\item Generate a point process $ (t_1, t_2, \ldots, t_{N^+})$ on the interval $(0, T)$ according to a simple Poisson process with rate $\lambda$. 
\item For each $j=1,\ldots, N^+$: 
\begin{description}
\item[(i)] Randomly choose a pair $(u_j, v_j)$ such that $ u_j \neq v_j$ from the population of $n$ vertices.
\item[(ii)] For vertices $u_j$ and $v_j$, generate $K$ -dimensional latent positions $X_{u_j}(t_j)$ and  $X_{v_j}(t_j)$ from distributions $F(\theta_{u_j}( t_j))$ and $F(\theta_{v_j}( t_j) )$ independently. The vertex-specific parameter processes $\theta_v(t)$ are described in section \ref{sec:cp}.
\item[(iii)] Unattributed edge case: generate a Bernoulli random variable $z_j$ such that 
\begin{equation}
P(z_j=1) = X_{u_j}(t_j)\cdot X_{v_j}(t_j)=\sum_{k=1}^K  x_{u_j}^{(k)}(t_j)x_{v_j}^{(k)}(t_j), 
\end{equation}  If $z_j =1$, draw an edge between $u_j$ and $v_j$. \\ Attributed edge case: Draw an element $k_j$ from the set ${0,1,2, \ldots, K}$ such that 
\begin{eqnarray}
\mathbf P( k_{j} = k ) = 
\left\{
  \begin{array}{ll}
    x_{u_j}^{(k)}(t_j) x_{v_j}^{(k)}(t_j) & \hbox{if } k > 0, \\
    1 - \sum_{k=1}^K  x_{u_j}^{(k)}(t_j)x_{v_j}^{(k)}(t_j) , & \hbox{if } k = 0,
  \end{array}
\right.
\end{eqnarray}
where  $x_{u_j}^{(k)}(t_j) $ and $ x_{v_j}^{(k)}(t_j)$ are the $k$th elements of vectors  $X_{u_j}(t_j)$ and  $X_{v_j}(t_j)$, respectively. If $k_j>0$, draw an edge with attribute $k_j$ between $u_j$ and $v_j$.

\end{description}

\end{enumerate}

The underlying  poisson process $(t_1, t_2\ldots, t_{N^+}) \in (0, T) $ creates edge `opportunities' uniformly over the graph with exponentially distributed inter-arrival times. For simplicity, we consider data observed over a fixed interval $(0, T)$, but a related  model for ongoing data acquisition could be constructed in which each new edge opportunity occurs at an exponentially distributed interval after the last.  Each pair of vertices has the same expected number of edge opportunities, and the expected number of edges opportunities does not change over time. The edge opportunity process is unobserved; what we observe are the realized edges,  a filtered version of the edge opportunity process. Through this filtering, we can model inhomogeneities in number and attributes of edges over time and over the graph.  A schematic for this two-level process is given in Figure 1. 
\begin{figure}[t]
  \centering
    \includegraphics[width=.8\textwidth]{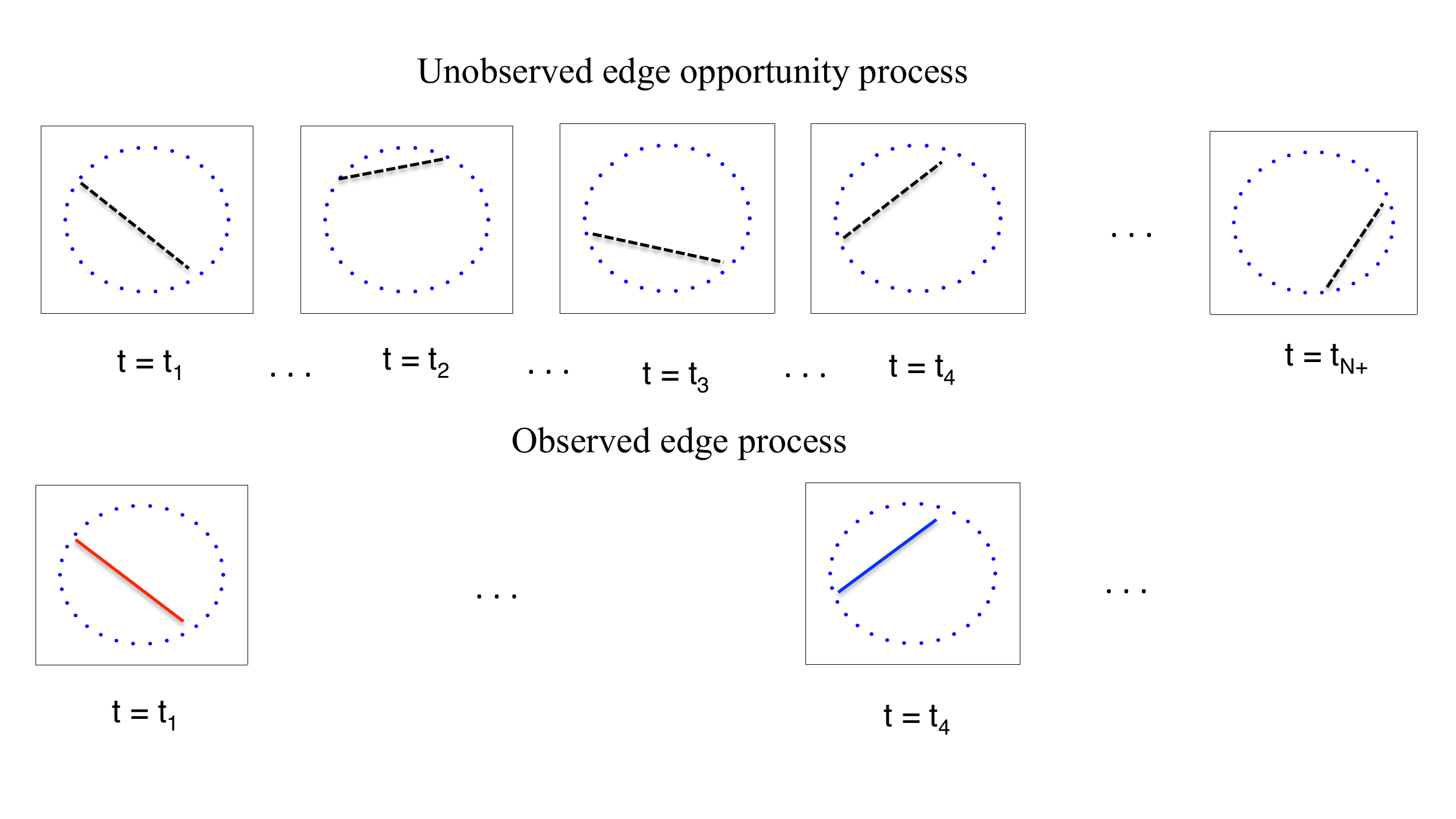}
    \caption{An underlying unobserved Poisson process (top) generates opportunities for edges between pairs of vertices. The observed edge process (bottom) is a filtered version of the edge opportunity process in which edges may have categorical attributes.  }
 \label{fig:edges}

\end{figure}

The probability an edge opportunity at time $t_j$  will be realized depends on the latent processes associated with vertices $u_j$ and $v_j$. At each time $t$, $u_j$ and $v_j$ occupy  positions $X_{u_j}(t)$ and $X_{v_j}(t)$  in a latent space. We define the latent space $\mathbb{S} $ to be the subset of $\mathbb{R}^K$ bounded by the $K$-dimensional simplex:
\begin{eqnarray}
\mathbb S = \{ x \in \mathbb R_+^{K} : \sum_{k=1}^{K} x_k \leq1\}.
\end{eqnarray}
The position of vertex $v$ in the latent space at time $t$, $X_v(t)$, is generated from  the distribution $F( \theta_v(t))$.  In what follows, we take $F$ to be the $(K+1)$-dimensional Dirichlet distribution. To allow for flexible modeling of edge probabilities, we take $X_v(t)$ to be the first $K$ components of a $(K+1)$-dimensional Dirichlet random variable, i.e. if $Y=(y_1,\ldots, y_{(K+1)})\sim \textrm{Dir}(\theta_v(t))$, $X_v(t) = (y_1, \ldots, y_K)$. Note that since Dirichlet random variables must sum to one, the $(K+1)^{\textrm{st}}$ component is fixed given the first $K$.  

The relative locations of the two vertices in latent space determine the probability that edges between them will be realized.  The probability an edge between vertices $u$ and  $v$ will be realized at time $t$ depends on $X_u(t)$ and $X_v(t)$ through the dot product:
\begin{equation}
P(z_j=1|X_{u_j}(t_j), X_{v_j}(t_j)) = P(k_j>0|X_{u_j}(t_j),X_{v_j}(t_j)) =  X_{u_j}(t_j) \cdot X_{v_j}(t_j) = \sum_{k=1}^K  x_{u_j}^{(k)}(t_j)x_{v_j}^{(k)}(t_j) .   
\end{equation}
The dot product captures inhomogeneities in the probability of connection across different vertex pairs. In particular, two types of inhomogeneities can be easily expressed through the dot product: differences in overall connectivity for an individual vertex,  and clustering, or the tendency for some groups of vertices to be more likely to be connected with one another. The former can be expressed through variations in the magnitude of the latent position vector. As $||X_v(t)||$ approaches 0 (\textit{i.e.}, the latent position approaches the origin), the dot product $X_v(t)\cdot X_u(t)$ will approach 0 for any other vertex $u$. Similarly, as $||X_v(t)||$ approaches 1, the probability of realized edges between $v$ and all other vertices will increase. A low probability of connection can also be expressed through  latent position vectors which are nearly orthogonal.

The second type of inhomogeneity, clustering, can be expressed through the angles between vertices. For vertex pairs with a large angle between them (different directions in the latent space), the dot product will be small, even if the overall magnitude of each vector is large. When vertex pairs point in the same direction in the latent space, they will be more likely to communicate.  

Using the dot product model, clusters in the latent space will give rise to clusters in the graph, as nearby vertices, with respect to angle from the origin, are more likely to communicate with one another. We are therefore interested in analyzing the structure of the graph in terms of distributions of latent positions.

The generative process described above produces a collection of events denoted $\mathbf{e}^+$ in the unattributed edge case and $\mathbf{a}^+$ in the  attributed edge case. Each event has a time stamp $t_j$, a vertex pair ($u_j, v_j$), and either an edge indicator variable $z_j$ or an attribute $k_j$:
\begin{equation}
\mathbf{e}^+ =  \{(t_j, u_j, v_j, z_j); j=1 \ldots N^+\}   = (T^+, U^+, V^+, Z^+) 
\end{equation}
and 
\begin{equation}
\mathbf{a}^+ = \{ (t_j, u_j, v_j, k_j); j=1 \ldots N^+\} =  (T^+, U^+, V^+, K^+)
\end{equation}
where $T^+ = \{t_j, j=1\ldots N^+\}$, $U^+ = \{u_j, j=1\ldots N^+\}$, $V^+ = \{v_j, j=1\ldots N^+\}$, $Z^+ = \{z_j, j=1\ldots N^+\}$, and $K^+ =\{k_j, j=1\ldots N^+\}$. 
The collections $\mathbf{e}^+$ and $\mathbf{a}^+$ contain events for which $z_j=0$ or $k_j=0$. Since no edges are created in these events, we do not observe them. We observe $\mathbf{e}$ or  $\mathbf{a}$ , subsets of $\mathbf{e}^+$ and  $\mathbf{a}^+ $for which edges are realized:
\begin{equation}
\mathbf{e} =  \{(t_j, u_j, v_j) : z_j=1\} =  \{(t_i, u_i, v_i) : i = 1 \ldots N\} = (T, U, V)
\end{equation}
 or
\begin{equation}
\mathbf{a} =  \{(t_j, u_j, v_j, k_j) : k_j>0\} = \{(t_i, u_i, v_i, k_i) : i = 1 \ldots N\} =(T, U, V, K). 
\end{equation}
As an example, consider \textbf{e} to be a record of email messages, each of which contains a sender/receiver pair and time label, and \textbf{a} to be a similar collection which also contains a categorical email topic attribute.  

Using the latent positions model, we are interested in performing inference on $\theta_1, \ldots ,\theta_n$, the parameters describing the distributions of the latent positions. If  $\theta_v =\theta$ for $v= 1 \ldots n$, each vertex pair will have the same expected number of edges. In the general case, if  $\theta_u = \theta_v$ for a particular pair $(u, v)$ vertices $u$ and $v$ may be more likely to communicate, depending on their overall probability of edge creation.  Given the latent positions, the probability of the complete collection of events can be factored  as

\begin{align}
 P( \mathbf{e}^+ |  \mathbf{X}, \lambda) 
   &=  P(U^+, V^+|N^+)P(Z^+|X, N^+)P(T^+, N^+|\lambda)\\
&  = C_{u, v, t}\prod_{j:z_j=1} X_{u_j}(\tau_j) \cdot X_{v_j}(\tau_j)\prod_{j:z_j=0}(1-X_{u_j}(\tau_j)\cdot X_{v_j}(\tau_j))  \, \frac{(\lambda T)^{N^+}e^{-\lambda T}}{N^+!}, \nonumber
  \end{align}
where $C_{ u, v, t}$ is a constant which does not depend on $X$ or $\theta$. The first product term describes the $N$ elements of \textbf{e}, i.e. the observed edges. The second term describes the subset of elements of \textbf{e}$^+$ which are unobserved, and therefore not in \textbf{e}. The number of edge opportunities $N^+$ and the latent vertex positions at unobserved events appear in the likelihood, but are unobserved. If we assume the Poisson process parameter $\lambda$ is given, as will be discussed in Section \ref{sec:fitting},  we can work with the expectation of the likelihood $l(\theta, \mathbf{e^+})$ given the observed data $\mathbf{e}$ :
 \begin{align}
E_{\mathbf{e}^+}(l(\theta, \mathbf{e}^+)|\mathbf{e}) &=
C_{ u, v, t}\sum_{N^+=0}^{\infty} \prod_{i=1}^N X_{u_i}(t_i) \cdot X_{v_i}(t_i)
 (1-E(X_u\cdot X_v|\theta))^{(N^+-N)}\frac{(\lambda T)^{N^+}e^{-\lambda T}}{N^+!}\\
& =  C_{ u, v, t}\prod_{i=1}^N X_{u_i}(\tau_i) \cdot X_{v_i}(\tau_i) \,  (1-E(X_u\cdot X_v | \theta))^{-N} e^{-( \lambda T) E(X_u\cdot X_v | \theta)},
\end{align}
where the second equality is simply an identity of the exponential function.  This allows us to fit the model based on the observed data $\mathbf{e}$ or $\mathbf{a}.$ 
\subsection{Change Points}
\label{sec:cp}

We are interested in detecting changes in the behavior of the network, particularly changes in which a subset of vertices alter their behavior as a group.  Where changes exist, we seek to partition the network over vertices and over time, detecting time-dependent inhomogeneities. Through an iterative procedure, it will ultimately be possible to detect multiple time points at which the network changes behavior, and multiple subpopulations of vertices exhibiting distinct behavior. As an initial step, we define a partition to identify unusual behavior in one subgroup of vertices over a single unknown time interval.  A schematic is given in Figure 2.


Recall that $X_v(t)$, the latent position of vertex $v$ at time $t$, is drawn from a distribution $F(\theta_v(t))$ . Changes in behavior for vertex $v$ through time are modeled through changes in $\theta_v$.  For the Dirichlet distribution, this is a vector  $\alpha = (\alpha_1, \ldots \alpha_{K+1})$, which specifies both the center and spread of the latent positions. Let $\mathbf{v} = \{v_1, \ldots, v_m\}$ be a subset of $\{1, \ldots, n\}$. Initially we compare two models:
\begin{align}
\textrm{Model 1}: \theta_v(t)  & = \theta_0, v \in \{ 1, \ldots n\} , t \in (0, T) \\
\textrm{Model 2}: \theta_v(t)  & = \theta_0, v \in \{1, \ldots, n\}, t \in (0, \tau_1) \cup (\tau_2, T), \label{hypotheses}
\\  \theta_v(t) &= \theta_0 , v \in \{1, \ldots, n\}/\mathbf{v}, t \in  ( \tau_1, \tau_2)  \nonumber \\
\theta_v(t) &= \theta_1, v \in \mathbf{v}, t \in ( \tau_1, \tau_2) \nonumber
\end{align}
for some unknown $ \theta_0, \theta_1, m$, $\mathbf{v} =\{ v_1, \ldots, v_m \}$, $\tau_1$ and $\tau_2$. Under model 1, the latent positions are drawn from the same distribution for all time and over all vertices. Under model 2, the latent positions of an unspecified group of vertices over an unspecified time interval are drawn from a different distribution, described by $\theta_1$. 
We approach the detection of multiple change points and multiple sub-populations of vertices in a hierarchical manner. An initial partition produces estimates of $(v_1, \ldots, v_m)$  and $(\tau_1, \tau_2)$ from the alternative model in (\ref{hypotheses}) using the EM algorithm, as described in the next section. If the initial partition is judged to fit the data significantly better than the homogeneous model, further partitions are evaluated.

\begin{figure}[t]
  \centering
    \includegraphics[width=.95\textwidth]{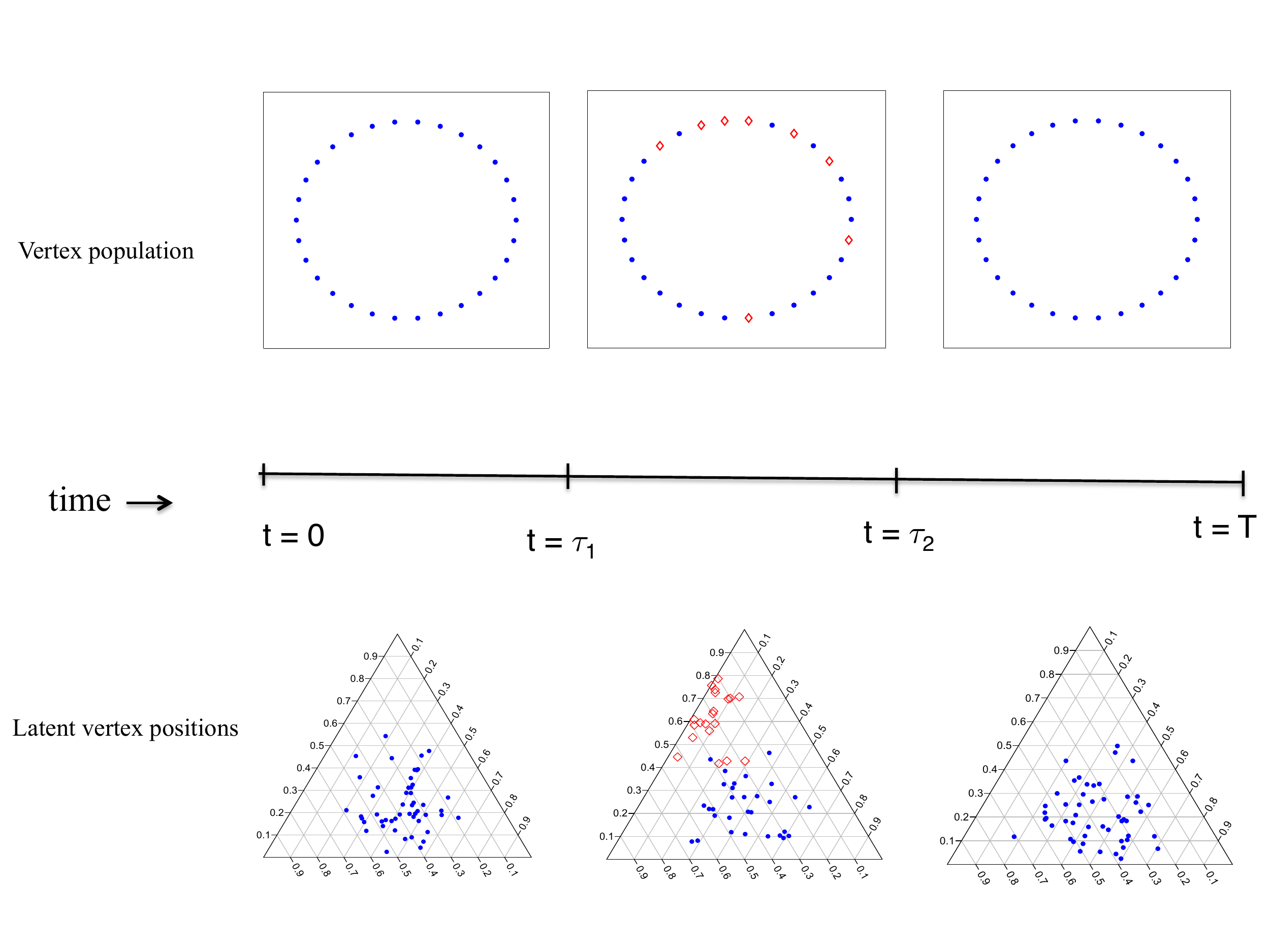}
    \caption{Over an unknown time interval ($\tau_1, \tau_2$), the population of vertices may contain distinct subpopulations with respect to latent positions in the $K$-dimensional simplex. In this case, K=2.}
\label{fig:vertchange}
\end{figure}

We seek to identify the number of partitions which optimally describes the data by defining a stopping rule to decide when further partitions are unnecessary. The hierarchical approach is used rather than simultaneously estimating multiple partitions over vertices and time because the latter is computationally impractical. 

Each partitioning stage  can be phrased in terms of the number of components in a mixture distribution. If there are no distinct subgroups in a given set of vertices, and behavior does not change over a given time interval, the latent positions can be modeled using one component distribution, \textit{i.e.},  $ X_v(t) \sim F(\theta_0), v \in \{ 1, \ldots n\} , t \in (0, T)$. If the distribution of the $X_v(t)$ differs over time or over vertices, then a mixture model with two or more components describes the distribution of the latent positions. We must then define a stopping rule which evaluates whether a two-component mixture model is a better description of  the vertices over a subinterval of  time  than a homogeneous model, according to some criteria. 

Determining the ``correct" number of components in a mixture distribution is a fundamental and arguably unresolved problem in cluster analysis.  Asymptotic results for test statistics  based based on modified likelihood ratios have been developed for simple parametric mixture models \citep{chenchenkalbfleisch}, which do not apply in our case. More general model-based approaches include AIC,  BIC \citep{schwarz1978} and the integrated completed likelihood (ICL, \citet{biernacki2000}) an entropy-based measure. Model selection criteria which do not make strong parametric assumptions have also been proposed, such as information based methods \citep{sugarjames2003}.
We find that BIC and ICL are the most reasonable model selection criteria in our case, and choose BIC over ICL for practical reasons, because the later is very computationally demanding in this setting. A simulation-based evaluation of the performance of the BIC-based decision rule is presented in Section \ref{sec:sims}.

\section{Model Fitting}
\label{sec:fitting}

A stochastic conditional EM algorithm \citep{mengrubin_ecm} is used to fit the model under heterogeneity.  The E-step at the $(j+1)^{\textrm{st}}$ iteration of the algorithm consists of computing $(\hat{\tau}_1, \hat{\tau}_2)^{j+1}$, a stochastic approximation to the conditional expectation of $(\tau_1, \tau_2)$ given the current parameter estimates  $\hat{\theta}_0^j$,  $\hat{\theta}^j_1$,  and $\mathbf{v}^j$ (the current estimate of $\mathbf{v} = \{v_1, \ldots v_m\}$), and an updated estimate $\mathbf{v}^{j+1}$ of $\mathbf{v}$. The M-step requires finding the updated maximum likelihood estimates $\hat{\theta}_0^{j+1}$ and $\hat{\theta}_1^{j+1}$ given the current estimates of $\tau_1, \tau_2 $ and $\mathbf{v}$. The non-standard aspects of the fitting algorithm are described in this section, with computational details available in the appendix. 

The success of the algorithm is dependent in part on the initial values used for the model parameters. Based on simulation results, the sensitivity of the algorithm to initial conditions seems to vary with the number of change points and of distinct subgroups of vertices, with more components increasing the sensitivity. An approach to producing initial values that we have found to work well is to divide  the entire observed time interval  $(0, T)$ into segments, find candidate values of $\mathbf{v} $ for each segment, and then assess which  combination of time interval and candidate $\mathbf{v}$ produces the highest likelihood under the 2-component mixture model. 

The details of the initial value procedure are as follows. The time interval $(0, T)$ is divided into $r$ equally sized segments, producing non-overlapping time intervals $(0, T/r), (T/r, 2T/r), \ldots ((r-1)T/r, T).$   In the unattributed edge case, the least squares estimates of the latent positions over each interval $\overline{\mathbf{X}}^1, \ldots, \overline{\mathbf{X}}^r$ are computed from the data in each time interval  using the SVD procedure described below.   This group of $n$  latent positions  are then clustered into two groups using k-means, producing an estimate of $\mathbf{v}$ for each interval.  In the attributed edge case, using the data from each interval,  a vector $(p_1, \ldots p_K)$ is created for each vertex, where $p_k$ is the proportion of edges with attribute $k$. These vectors are then clustered using c-means, producing estimates of  $\mathbf{v}$ for each interval.

The E-step at the $(j+1)^{\textrm{th}}$ iteration of the algorithm consists of computing a stochastic approximation to the conditional expectation of $(\tau_1, \tau_2)$ given the current parameter estimates  $\theta_0^j$,  $\theta^j_1$,  and $\{v_1, \ldots v_m\}^j$ and the current estimates of $\mathbf{v}$.  The approximate expectation is produced by simulating a large number (5,000, in our case) of 
candidate intervals $\{(t_{1, z}, t_{2,z}): z=1, \ldots 5000\}$ from the uniform distribution over $(0, T) \times (0, T)$ and computing  
\begin{equation} 
p(t_{1, z}, t_{2,z} | \mathbf{e} , \mathbf{v}^j, \theta_0^j, \theta_1^j) = \frac{p(\mathbf{e}|t_{1, z}, t_{2,z}, \mathbf{v}^j, \theta_0^j, \theta_1^j)\frac{1}{Z}}
{\frac{1}{Z}\sum_{z=1}^Z p(\mathbf{e}|t_{1, z}, t_{2,z}, \mathbf{v}^j, \theta_0^j, \theta_1^j)}  
\end{equation}
for each interval, where $p(\mathbf{e}|\ldots)$ is calculated using (\ref{likelihood_at}) or (\ref{likelihood_unat}). The collection of intervals and their associated conditional probabilities are then use to compute the approximate expectation $(\hat{\tau}_1, \hat{\tau}_2)^{j+1}$. 

The new estimate $\mathbf{v}^{j+1}$ is then computed using the updated estimate  $(\hat{\tau}_1, \hat{\tau}_2)^{j+1}$. For $i=1, \ldots,n$,  we estimate the probability that vertex $i$ is in the anomalous group $\mathbf{v}$ over the interval  $(\hat{\tau}_1, \hat{\tau}_2)^{j+1}$ given  the current parameter estimates and the current estimates of the other members of the anomalous group. We will assign vertex $i$ to the updated estimate  $\mathbf{v}^{j+1}$ if
\begin{eqnarray}
p(i \in \mathbf{v}| \mathbf{e}, (\hat{\tau}_1, \hat{\tau}_2)^{j+1}, \hat{\theta}_0^{j},  \hat{\theta}_1^{j}, \mathbf{v}^j)>\xi
\end{eqnarray}
for some threshold $\xi$, where we use $\xi=.5$. The condition can be evaluated via Bayes rule and (\ref{likelihood_at}) or (\ref{likelihood_unat}).

Estimates $\hat{\theta}_0^{j+1}$ and $\hat{\theta}_1^{j+1}$ given the current estimates of $\tau_1, \tau_2 $ and $\mathbf{v}$ are computed in the M-step of the algorithm.  This is somewhat complex as it involves integrating over the unobserved latent positions as well as dealing with non-identifiability issues. As in other latent position models \citep{handcockraftery2007,gormleymurphy2010}, the likelihood in the unattributed edge dot product model is invariant to reflections and rotations of the latent positions (although it is not invariant to translations.)  The likelihood is convex with respect to the dot product of the latent positions, but not necessarily with respect to the latent positions themselves.  In the attributed edge case, the likelihood is not invariant to rotations and reflections, because the directions in the latent space have meaning with respect to the probabilities of various edge attributes. \citet{handcockraftery2007} resolve the non-identifiability in the latent distance model by post-processing the MCMC output to find the configuration of latent positions which is optimal in terms of Bayes risk.

In order to resolve the non-identifiability for the unattributed random dot product model, we define an additional condition based on minimizing the squared difference between a modified adjacency matrix and a matrix consisting of the expected number of edges between each pair. 

Let $\overline{X}_v$ be the average latent position of vertex $v$ over a given time interval $(t_1, t_2)$. The collection  of latent positions over all vertices is the $K \times n $ matrix  $\overline{\mathbf{X}} = ( \overline{X}_1, \ldots \overline{X}_n)$. The expected number of edges over $(t_1, t_2)$ between vertices $u$ and $v$ is the $(u, v)^{\textrm{th}}$ element of the matrix  
$b\overline{\mathbf{X}}^T \overline{\mathbf{X}}$, where
\begin{equation}
 b= \frac{\lambda(t_2-t_1)}{ {n\choose 2} }.
\end{equation}
Let $A$ be the multiadjacency matrix consisting of the number of edges between each vertex pair over the interval $(t_1, t_2)$.  We would like to compute  an initial estimate of the latent positions by minimizing the difference between the observed and expected number of edges for each pair, given the latent positions. We cannot directly compare $A$ and $B\overline{\mathbf{X}}^T \overline{\mathbf{X}}$, however, because the diagonal entries in the two matrices are not comparable. We instead use  $\tilde{A}$, a modification of  $A$ with an augmented diagonal as described in \citet{scheinermantucker}. Using singular value decomposition, we can easily find the latent positions $\overline{\mathbf{X}}$ which minimize

\begin{equation}
\|b\overline{\mathbf{X}}^T \overline{\mathbf{X}} - \tilde{A}\|_F^2,
\end{equation}
 the squared Frobenious norm of the difference between expected and observed edges.
This least squares estimate of the latent positions is used as a starting point to a hill-climbing procedure to find the local maximum likelihood estimates of $\theta, \theta_0 $ and $ \theta_1$ which are closest to the minimizer of the least-squares condition. The computational details of the fitting algorithm are given in the appendix.


In our model, there is non-identifiability between the Poisson process rate $\lambda$ and the parameters of the latent position distribution, $\theta$ which can be easily resolved.  The expected number edges created in a given time interval increases with both $\lambda$  and the expected value of $X_u\cdot X_v$; increasing $\lambda$ produces more edge opportunities, and increasing $E(X_u \cdot X_v )$ increases the probability of an edge at each opportunity. We approach this non-identifiability by fixing $\lambda$ to be 1.5 times the maximum number of edges observed in any time unit (weeks, in our application). Thus the expected number of edge opportunities (not necessarily edges) in a unit time interval is 50\% greater than the maximum observed  number.

\section{Simulations}
\label{sec:sims}
 We simulate  network datasets   \textbf{a}  = $\{a_1, \ldots, a_N\}$ using the generative model in Section \ref{sec:model} to test the performance of the partitioning algorithm. At random times  $t_1, \ldots, t_{N^+}$, pairs of latent positions \\
 $\{(X_{u_1}, X_{v_1}), \ldots ,( X_{u_{N^+}}, X_{v_{N^+}})\}$  are drawn from the $(K+1)$-dimensional Dirichlet distribution. Change points are inserted at times $\tau_1$ and $\tau_2$, between which a subset of vertices $\{v_1, \ldots v_m\} \in \{1, \ldots, n \}$ has latent positions drawn from a Dirichlet distribution with parameter $\alpha_2$ rather than $\alpha_1$.  

\begin{figure}[t]
  \centering
    \includegraphics[width=.6\textwidth]{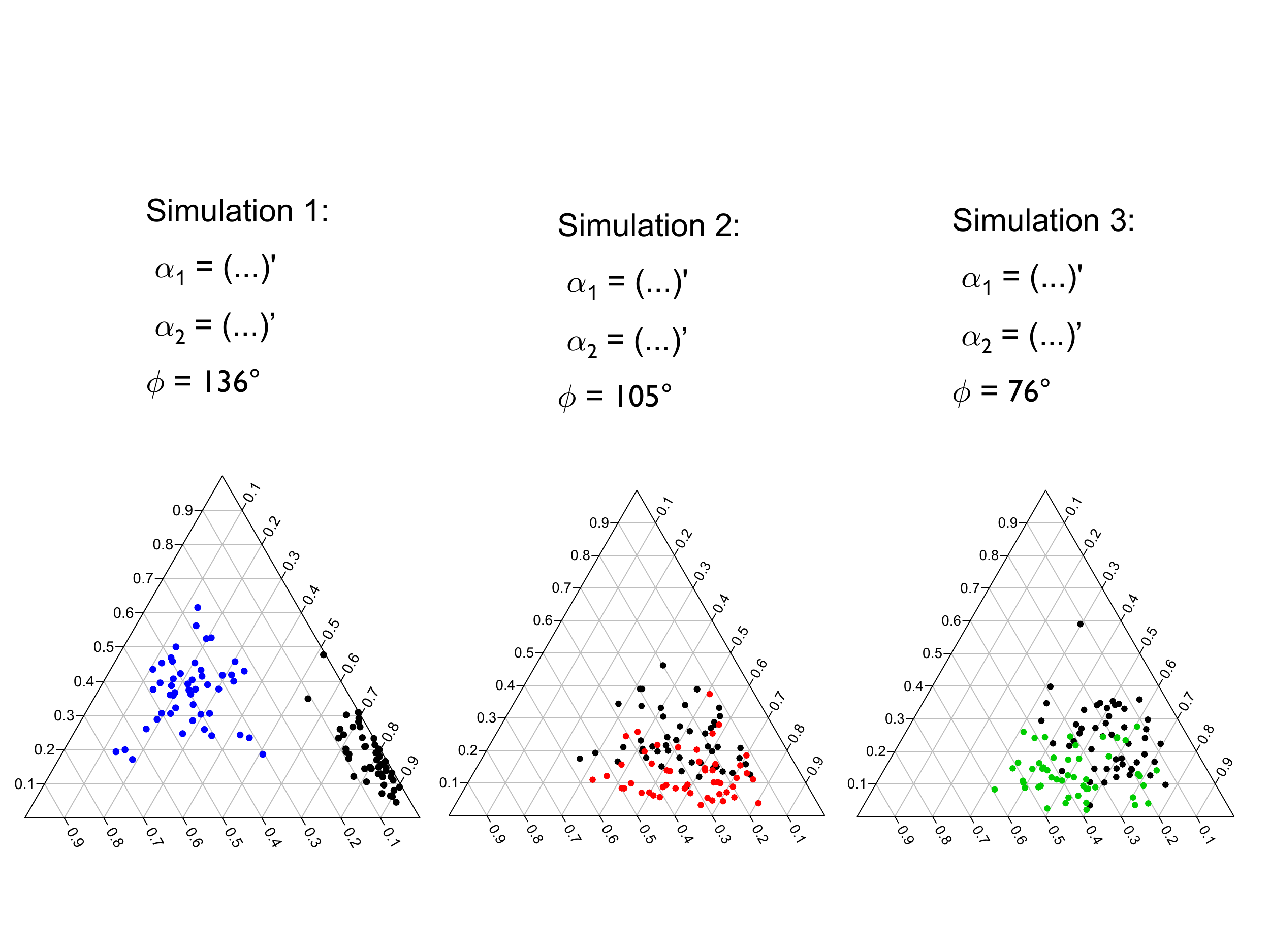}
    \caption{Distributions of the latent positions used to create simulated network data. The Dirichlet parameters $\alpha_1$ (non-anomalous subset) and $\alpha_2$ (anomalous subset are shown,) along with the angle $\phi$ between $\alpha_1$ ad $\alpha_2$ and an example of 50 simulated latent positions from each distribution (bottom). Latent positions are shown in a 3-D projection.  }
 \label{fig:simsalpha}
\end{figure}

Performance is evaluated under different possible network characteristics by simulating datasets with varying values of the Dirichlet paramters ($\alpha_1, \alpha_2$), number of vertices \textit{n}, and and poisson process rate $\lambda$.   We consider three possible $(\alpha_1, \alpha_2)$ combinations. The distance between the centroids is summarized by $\phi$, the angle between them, which is proportional to the overall probability of connection between the anomalous and non-anomalous vertices. Network characteristics also depend on the spread of the Dirichlet distributions,  with higher variance leading to decreased edge density overall and a higher tendency for edges to cluster.  The three simulations scenarios, along with illustrative examples of random draws from each latent position distribution,  are shown in Figure \ref{fig:simsalpha}. We simulate data for $n=50$ and $n=150$ vertices over a total time interval (0, 100), with varying densities of edges over time, as controlled by the Poisson process rate  $\lambda$. The length of the anomalous interval is 40\% of the total observed interval. In each simulation, the number of anomoulous vertices is $m=10$, representing 20\% of all vertices when $n=50$ and 6.7\% of vertices when $n=150$.

  The ability to detect time periods with heterogeneous subgroups increases with $\phi$, the angle between $\alpha_1$ and $\alpha_2$, $\overline{N} = N/(n(n-1)/2)$, the average number of edges per pair, $\tau_2-\tau_1$, the length of the anomalous interval,  and with $m$, the size of the anomalous subgroup (up to $m=n/2$).  The average number of edges per pair is a function of the length of the total time interval, the number of vertices $n$, the Poisson process rate $\lambda$ and the values of $\alpha_1$ and $\alpha_2$.  We use four measures to assess how well the fitting algorithm and BIC-based decision rule perform in each simulation condition.  The detection power, as measured by the number of times that the homogeneous model is rejected in favor of the heterogenous model based on BIC, is shown in Figure \ref{fig:simsalpha}.   Performance is also measured by the correct identification of vertices as anomalous and non-anomalous in simulations where the homogeneous model is rejected.  The sensitivity (correct identification of anomalous vertices) and specificity (correct identification of non-anomalous vertices) are shown in Figure \ref{fig:sims},  as well as   $  (|\tau_1 - \hat{\tau}_1| + |\tau_2 - \hat{\tau}_2 |)/2$, the average error in estimates of the change points $\tau_1$ and $\tau_2$, when they are detected.
  Because of the invariance discussed in Section \ref{sec:fitting}, recovery of  $\alpha_1$ and $\alpha_2$  is not considered.

   \begin{figure}[t]
\label{fig:powersims}
  \centering
    \includegraphics[width=.6\textwidth]{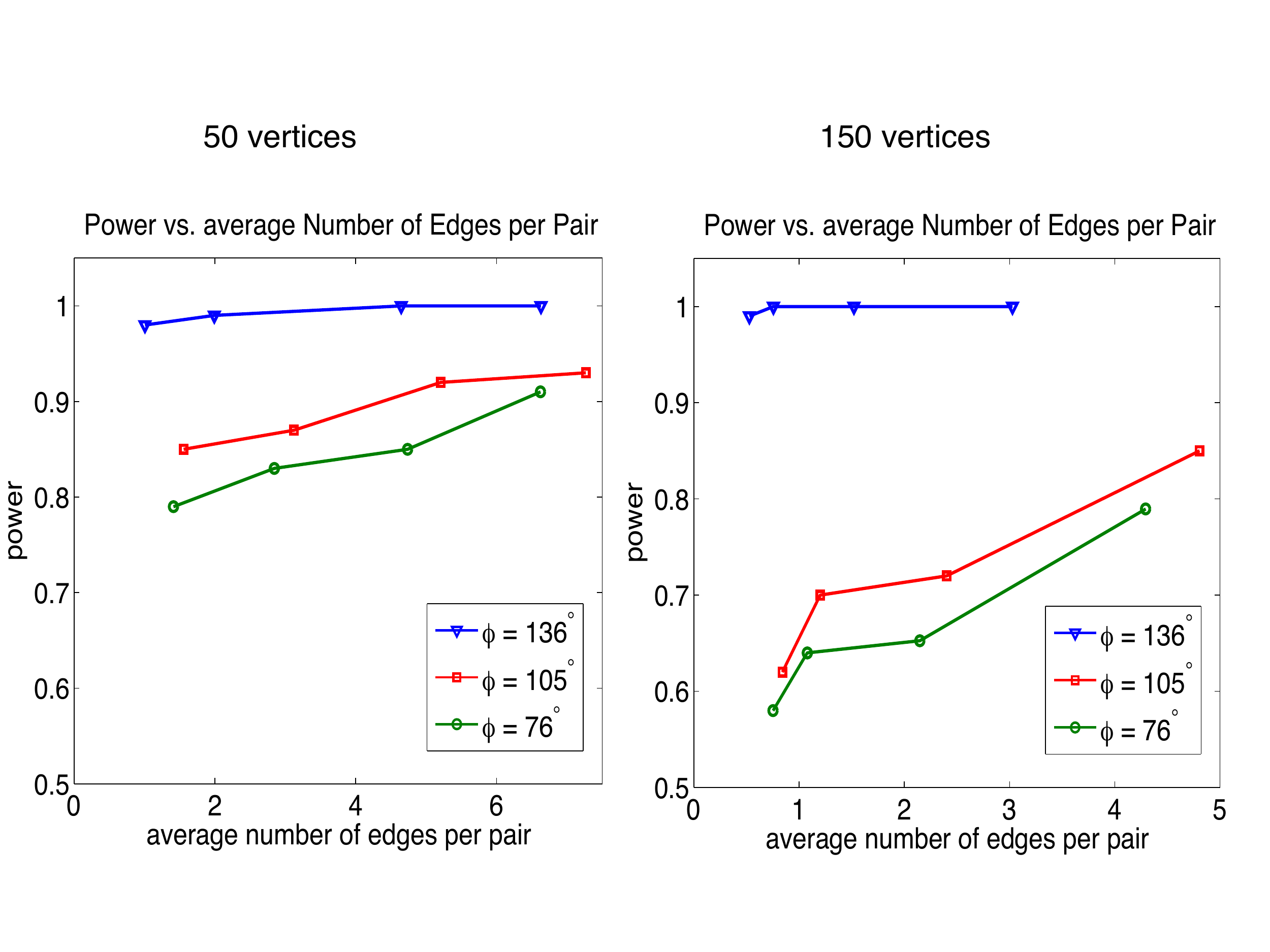}
  \caption{Power results for simulations. The observed probability of rejecting the homogeneous model in favor of the heterogeneous model using BIC vs average number of edges per pair is shown for the three simulation scenarios depicted in Figure \ref{fig:simsalpha} and n=50 (left) and n=150 (right), holding the number of anomalous vertices constant at m=10.   }
\end{figure}
    
   \begin{figure}[t]
  \centering
    \includegraphics[width=\textwidth]{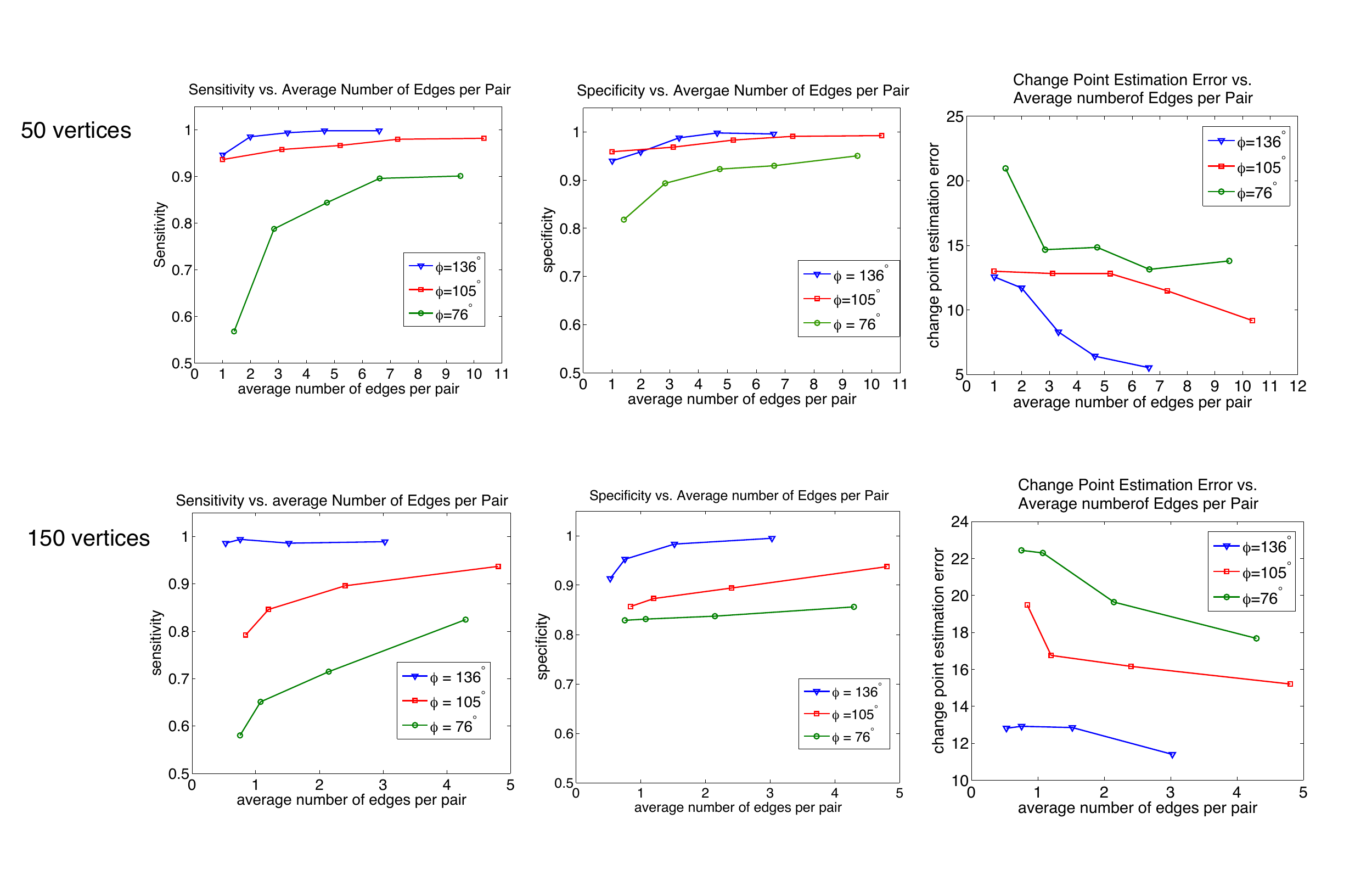}
  \caption{Simulation results for the change point detection algorithm. The figures show sensitivity in detecting anomalous vertices (left), specificity (center), and average change point detection error  (right). Results based on 100 simulations are shown for n=50 (top row) and n=150 (bottom row).  }
\label{fig:sims}
\end{figure}

In the simulated data with the smallest difference between the anomalous and non-anomalous latent positions (simulation 3 in  Figure \ref{fig:simsalpha}), power in detecting anomalous sub-groups exceeds 80\% in networks with 50 vertices and  average number of edges per-pair of 2. Power increases to approximately 90\% as the average number of edges increases to 6. Specificity  follows a similar pattern, and sensitivity is above 80\% for average per-pair degree greater than 1. Performance improves with larger differences between subsets. 
When the number of vertices increases to 150 and the percentage of anomalous vertices drops to 6.7\%,  performance decreases, particularly for in the case of the smallest separation between subgroups.  
In the least-separated simulations, power in detecting anomalous sub-groups exceeds 60\% in for networks with  an average degree of 1.1. Power increases to approximately 80\% as the average number of edges increases to 5. Sensitivity follows a similar pattern, and specificity is above 80\% for average per-pair degree greater than 1. Performance improves with larger differences between subsets, with values above 85\% and 90\%, and above 95\% for the moderately and highly separated distributions respectively.

\section{Application to Enron Data}

As an application, we analyze the Enron email corpus \citep{enronsite}. Following its investigation of Enron, the Federal Energy Regulatory Commission publicly released emails sent between 1998 and 2002  among 184 email addresses belonging to roughly 150 Enron employees, including high-level executives. We a use version of the dataset which has been processed to correct some integrity problems \citep{priebescanstats}. A Ring plot of the Enron data is shown in Figure \ref{fig:enronring}. Previous analyses \citep{fusongxing, diesnerenron, priebescanstats} have found these data to  exhibit heterogeneity over time, and to exhibit structure across vertices. In order to explore heterogeneity with respect to attributed edges, we use Michael Berry's topic classifications for the 2001 data \citep{ldctopics}, which assigns one of 32 topics to most messages. Using the assigned topics as edge attributes, we apply our algorithm to study time-dependent  structure in the dataset over the calendar year 2001.  

   \begin{figure}[t]
  \centering
    \includegraphics[width=.6\textwidth]{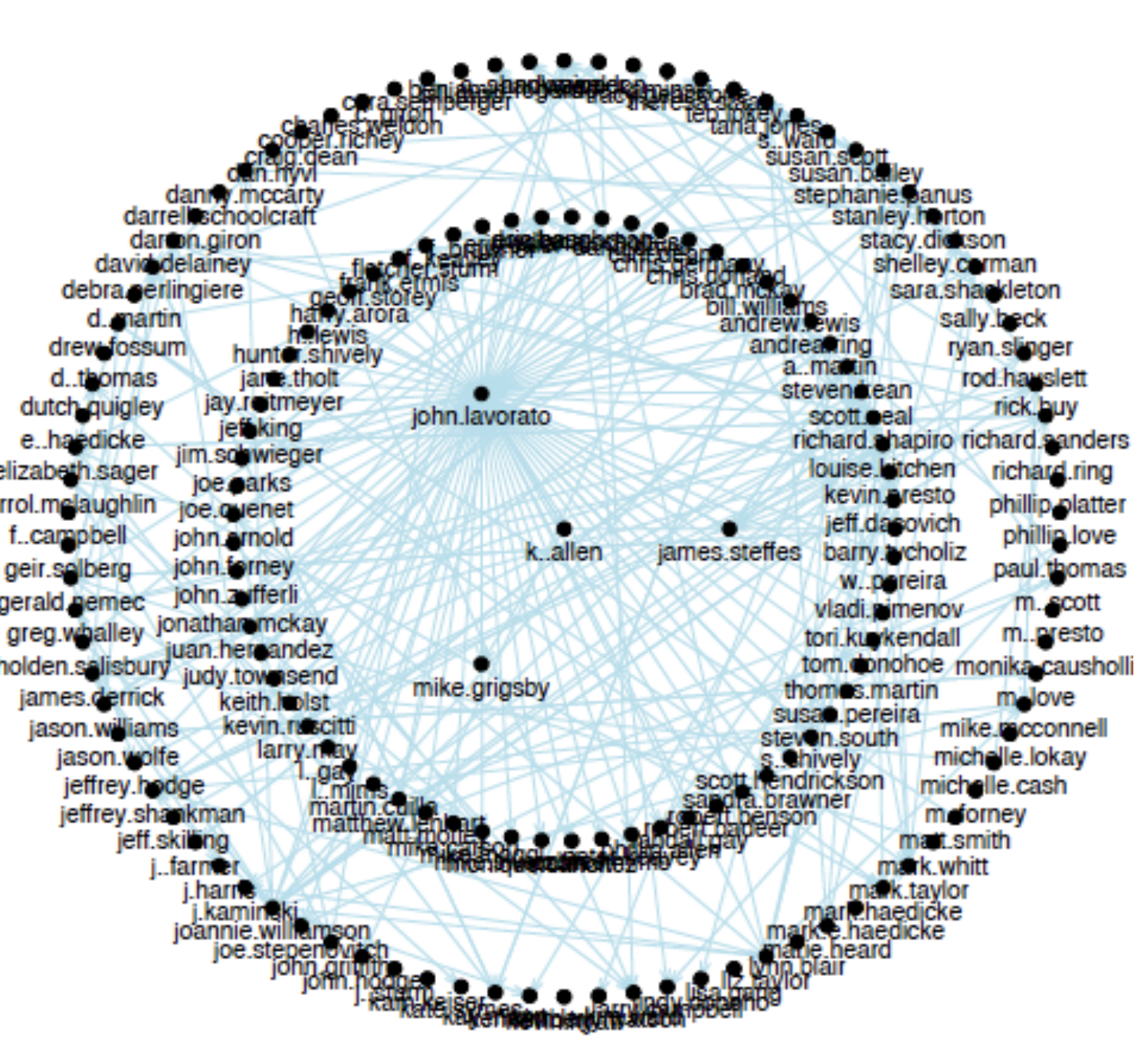}
  \caption{Email connections between enron emplyees in 2001.  }
\label{fig:enronring}
\end{figure}

 \citet{fusongxing} have  analyzed these data using the dynamic mixed-membership stochastic block model. Using information on the direction of the connections, the blockmodel provides a useful categorization of email behavior.  Fu et al. identify, for example, groups of employees who tend to send and receive emails among themselves as cliques,  groups that receive emails but rarely send them, etc.  Their analysis of the changing membership vectors associated with individual email addresses reveals changes in the frequency of emails in late 2001 as the company was going bankrupt, and a decrease in emails sent by high-level executives over the same period.

Analysis of the Enron data using the attributed edge dot product model reveals multiple change points. We restrict our attention to data on emails which have been assigned one of the 4 most popular email topics over the time interval:  business related to California , energy trading, general daily business, collapse of the company, and the Federal Energy Regulatory Commission and Department of Energy.The average number of connections between pairs is 1.1 Although this is a subset of the entire email corpus, and likely contains some residual integrity problems, we find it to be an informative application for our partitioning algorithm.  

In the attributed model, the dominant change point detected (that which is identified in the first partition) is in mid-August 2001, after which the detected heterogeneity persists until mid-December 2001. This time interval is significant with respect to major events related to the collapse of the company. CEO Jeffrey Skilling resigned on August 15, 2001 and over the following weeks the Enron stock price decline sharply and concerns over the company's accounting practices became public. Enron declared bankruptcy in December 2001. The overall message rate increased somewhat in this period with respect to the January - August period (423 vs 297 messages per week, on average.) However, the graph partitioning algorithm identifies a subgroup of 36 email addresses for which communication decreases significantly. There is also a shift in distribution of message subjects (edge attributes). A second anomalous time period is detected between mid April and late July. Here, a group of 88 email addresses show a greatly increased communication rate and a change in email attributes. A schematic summary of the detected change points is given in Figure 7. These results are consistent with the ``chatter anomalies"  detected in \citet{priebescanstats} using scan statistics on an unattributed version of the same data. 
   \begin{figure}[t]
  \centering
    \includegraphics[width=.4\textwidth]{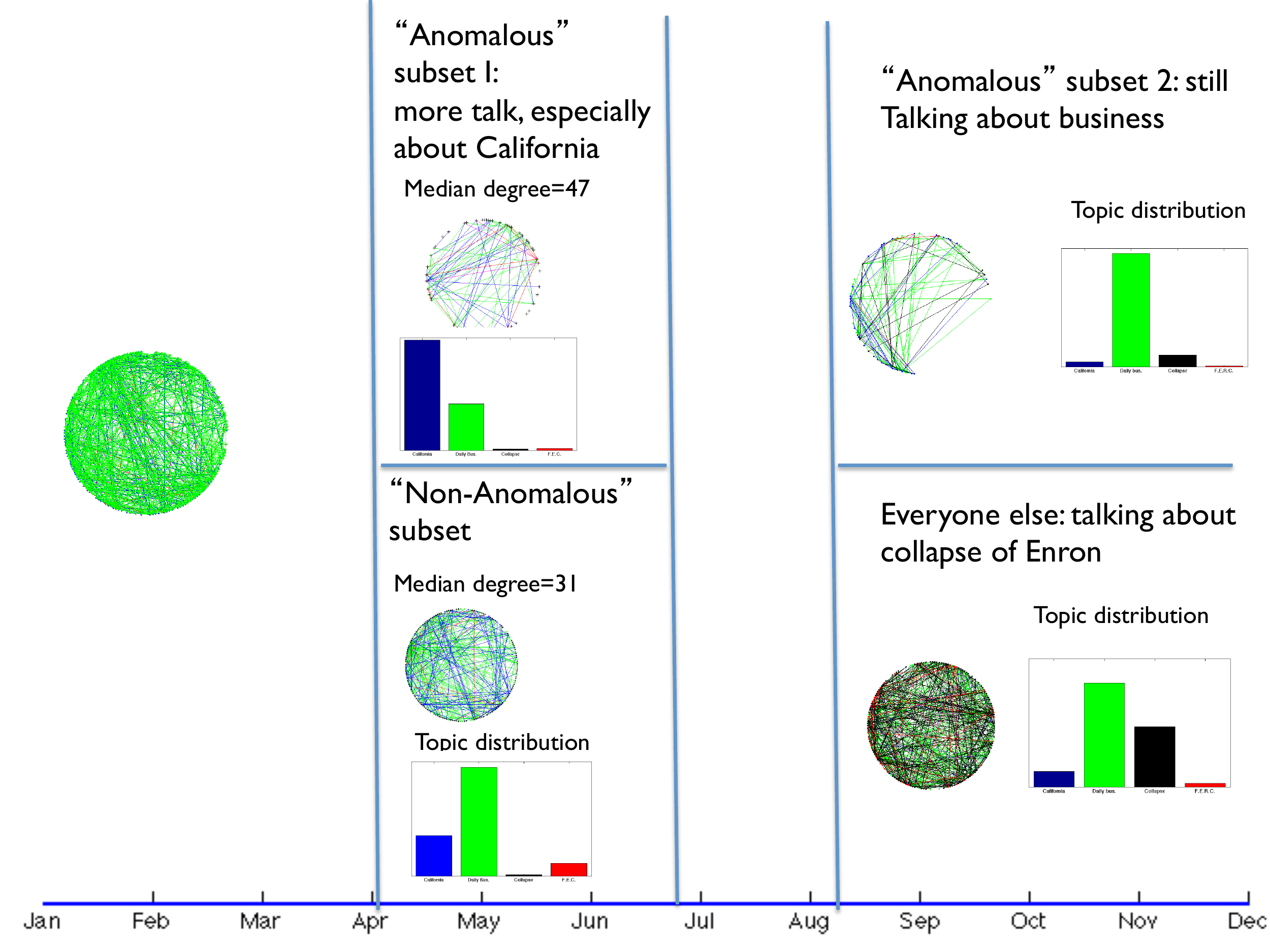}
  \caption{Email connections between enron emplyees in 2001.  }
\label{fig:enronring}
\end{figure}

There are several features of the Enron data that the latent process model does not reflect. In particular, there is considerable skew in the degree distribution of the email addresses, with some individuals exhibiting much higher connectivity than would be predicted by a model with homogeneous underlying Poisson edge creation process. This phenomena is largely explained by individuals who regularly send mass emails as opposed to more personalized communication.  The direction (sender to receiver) of the of the emails is also not addressed. The edge creation model could be modified to reflect directed connections or the long-tailed degree distribution within the Poisson process framework. Despite the aspects of the data which are not well-modeled in this analysis, we find this method to be a useful exploratory tool in identifying subgroups of individuals and time sub-intervals which exhibit distinct and interesting behavior.

\section{Discussion}

We have presented a novel algorithm and associated model for time-dependent network partitioning capable of discovering dynamic network structure. An important feature of this model is the inclusion of potential edge attributes, facilitating a more rich description of datasets such as the Enron email corpus. The filtered Poisson process model is a flexible framework for describing inhomogeneities in the rate of edge creation over time and over the graph.  The detection of meaningful partitions is approached by fitting a mixture model  and comparing it to a model with a homogeneous distribution of latent positions.

The model fitting algorithm simultaneously estimates latent positions and cluster structure, which has been shown in related contexts \citep{handcockraftery2007} to give better results than a two-stage procedure (position estimation followed by clustering). We approach the detection of multiple change points and multiple vertex subpopulations by iterating the partitioning algorithm.  Iterative partitioning may not perform as well as a  simultaneous estimation of multiple components, but we find the latter to be computationally infeasible. In order to test each partition for significance, we use the modified likelihood ratio test statistic proposed by \citet{chenchenkalbfleisch}.
Fitting the random dot product mixture model presents some computational challenges. Like the latent-distance model \citep{hoffrafteryhandcock}, the likelihood surface described by the random dot product model is non-convex, and the latent positions are non-unique. We impose an additional condition based on the least-squares optimal latent positions in order to define a unique maximum likelihood solution. We find this to be a fast and reliable solution to the non-uniqueness problem, although the precise relationship between the least-squares optimal solution estimates and the maxima of the likelihood surface  is unclear.

\citet{josephwolfson} discuss the consistency of  maximum likelihood estimates in the ``multi-path" change point problem, in which a collection of time series are observed, each potentially containing a unique change point.  The locations of change points in individual time series are assumed to be independent of one another.  They find that under certain regularity conditions, the maximum likelihood estimates of the change point locations and of the pre-and post-change parameters are consistent estimators. In our context, each vertex has an associated time course with a possible change point. In contrast to the case described by \citet{josephwolfson}, the change points of individual vertex time courses are not independent of one another; this non-independence makes analytic results on the asymptotic behavior of the parameter estimates difficult. 

The proposed model for edge creation has a simple functional form, and while it is able to capture features frequently observed in real network data, it may not incorporate all available information.  Several extensions of the model are possible. In particular, we have allowed for the parameters distributions of latent positions to change in time, but have not incorporated temporal dependence in the latent processes associated with each vertex. \citet{leepriebe2010} have analyzed the performance of inferences based on first and second order approximations to time-dependent latent dot product models. It may be possible to (for example)  fit an autoregressive component to the proposed latent process model by updating latent position estimates with each new edge event, conditional on the estimated positions  at the last edge event. For data in which the overall rate of edges is known to change over time, the model could be modified in a simple manner by specifying a non-constant  Poisson process rate $\lambda$. Under such conditions, the partitioning algorithm would detect changes in network behavior with respect to the overall dynamic message rate.  In some applications, for example the Enron email data, vertex-level covariates and directional information are also available. A model which includes covariate information could be used in place of the simple dot product model.
In general, the filtered Poisson model and change-point framework described above may be useful in combination with other latent position models for network data.

\appendix
\section{Appendix}
Here, we present more detailed information on the fitting algorithm for both the attributed and unattributed edge cases. 

\subsection{Attributed edge case} 

Let $y_i$ and $z_i, i=1, \ldots N$ be indicators of whether the vertices associated with each edge event are  contained in the set $\mathbf{v}$, and  $s_i, i=1, \ldots N$ be indicators of which edges occur during the interval $(\tau_1, \tau_2)$: 
\begin{eqnarray} \label{indicators}
 y_i = 
\left\{
  \begin{array}{ll}
    1 & \hbox{if } u_i \in  \mathbf{v}, \\
    0 &  \textrm{ otherwise}
  \end{array}
\right. ,
 z_i =  
\left\{
  \begin{array}{ll}
    1 & \hbox{if } v_i \in \mathbf{v}, \\
    0 &  \textrm{otherwise},
  \end{array}
\right.
 \quad i=1, \ldots N,
 \\
 s_i =  
\left\{
  \begin{array}{ll}
    1 & \hbox{if } t_i \in (\tau_1, \tau_2), \\
    0 &  \textrm{otherwise},
  \end{array} \nonumber
\right.
 \quad i=1, \ldots N.
\end{eqnarray}
At each edge event  $a_i = (t_i, u_i, v_i, k_i)$, the associated latent positions $X_{u_i}$ and $X_{v_i}$ are each drawn from either $f(\alpha_1)$ or $f(\alpha_0)$, where $f(\alpha)$ is the Dirichlet distribution, depending on the values of $\tau_1, \tau_2, $ and $\mathbf{v}$. We can classify each edge event according to whether 0, 1 or 2  of the associated vertices are drawn from $f(\alpha_1)$ rather than $f(\alpha_0)$.  Let $N_0, N_1, $ and $N_2$ be vectors of length $K$ containing the frequency of edge attributes for for events containing 0, 1, or 2 latent positions drawn from $f(\alpha_1)$. The  $k$th respective elements  of these vectors are  
\begin{align}
N_{0, k} = \sum_{i=1}^N ((1-s_i) + s_i(1-y_i)(1-z_i))* \mathbf{1}_{k_i = k}, \\
N_{1, k} = \sum_{i=1}^N (s_i y_i (1-z_i) +s_i z_i(1-y_i))*\mathbf{1}_{k_i = k}, \\
N_{2, k} = \sum_{i=1}^N s_i y_i z_i*\mathbf{1}_{k_i = k}, \\
 k=1, \ldots K.
\end{align}
Let
\begin{align}
\overline{N}_0 = \sum_{k=1}^K N_{0, k}, \quad \overline{N}_1 = \sum_{k=1}^K N_{1, k}, \quad \overline{N}_2 = \sum_{k=1}^K N_{2, k}, 
\end{align}
and
\begin{align}
\gamma_0 = \lambda (T - (t_2 - t_1)\frac{ \binom{n-m}{2}}{\binom{n}{2}} ), \quad \gamma_2 = \lambda(t_2-t_1)\frac{\binom{m}{2}} { \binom{n}{2}}. 
 \quad \gamma_1 = \lambda T - \gamma_0 - \gamma_2,
\end{align}
where $m$ is the number of elements in the set $\mathbf{v}$. 
The E-step requires computing the likelihood of the data given the model parameters as a  function of the Dirichlet parameters $\alpha_0 = (\alpha_{0, 1}, \ldots, \alpha_{0, K+1})'$ and $\alpha_1 = (\alpha_{1, 1}, \ldots, \alpha_{1, K+1})'$ after integrating over the unobserved latent positions:
\begin{align}
 \label{likelihood_at}
p(\mathbf{e}| \alpha_0, \alpha_1, \tau_1, \tau_2, \mathbf{v}) \propto \prod_{i=1}^N\frac{\alpha_{s_iy_i, k_i}\alpha_{s_iz_i, k}}{\overline{\alpha}_{s_iy_i}\overline{\alpha}_{s_iz_i}}
\Big{(}1-\sum_{k=1}^K \Big{(}\frac{\alpha_{0, k}}{\overline{\alpha}_0}\Big{)}^2\Big{)}^{(\gamma_0-N_0)}\\
\times
\Big{(}1-\sum_{k=1}^K \frac{\alpha_{0, k}\alpha_{1, k}}{\overline{\alpha}_0\overline{\alpha}_1}   \Big{)}^{( \gamma_1-N_1)} \nonumber
\Big{(}1-\sum_{k=1}^K \Big{(}\frac{\alpha_{1, k}}{\overline{\alpha}_1}\Big{)}^2\Big{)}^{(\gamma_2-N_2)},
\end{align}
where
\begin{align}
\overline{\alpha}_0 = \sum_{k=1}^{K+1} \alpha_{0, k}, \quad
\overline{\alpha}_1 = \sum_{k=1}^{K+1} \alpha_{1, k}.\\
\end{align}
The CM-step of the algorithm is to compute updated estimates $\hat{\alpha}_0^{j+1}$ and $\hat{\alpha}_1^{j+1}$ based on the updated estimates of $\hat{\tau}_1^{j+1},\hat{ \tau}_ 2^{j+1}$ and $\mathbf{v^{j+1}}$ and the estimates  $\hat{\alpha}_0^j$ and $\hat{\alpha}^j_1$ from the last iteration.  Let $s_i, y_i,$ and $ z_i$ be defined as in (\ref{indicators}) using $\hat{\tau}_1^{j+1},\hat{ \tau}_ 2^{j+1}$ and $\mathbf{v}^{j+1}$. Then

\begin{align}
\hat{\alpha}_{0, k}^{j+1} = \frac{ \overline{\alpha}_0^j\{((\gamma_1\hat{\alpha}_{1, k}^j/\overline{\alpha}_1^j )^2+
8\gamma_0(N_{0, k}+N_{1, k}))^{1/2}-\gamma_1\hat{\alpha}_{1,k}^j/\overline{\alpha}_1^j\}}
{2\gamma_0}, \quad k=1, \ldots, K, 
\end{align}
\begin{align}
\hat{\alpha}^{j+1}_{0, K+1} =\overline{\alpha}_0^j\Big{(}\frac{(\sum_{k=1}^K \hat{\alpha}^{j+1}_{0,k})^{1/2}}{\frac{\overline{N}_0}{\gamma_0}} - \sum_{k=1}^{K} \hat{\alpha}_{0, k}^{j+1} \Big{)}, 
\end{align}

\begin{align}
\hat{\alpha}_{1, k}^{j+1} = \frac{ \overline{\alpha}_1^j\{((\gamma_1\hat{\alpha}_{0, k}^j/\overline{\alpha}_0^j )^2+
8\gamma_2(N_{2, k}+N_{1, k}))^{1/2}-\gamma_1\hat{\alpha}_{0,k}^j/\overline{\alpha}_0^j\}}
{2\gamma_2}, \quad k=1, \ldots, K, 
\end{align}
\begin{align}
\hat{\alpha}^{j=1}_{1, K+1} =\overline{\alpha}_1^j\Big{(}\frac{(\sum_{k=1}^K \hat{\alpha}^{j+1}_{1,k})^{1/2}}{\frac{\overline{N}_1}{\gamma_1}} - \sum_{k=1}^{K}\hat{ \alpha}_{1, k}^{j+1} \Big{)}.
\end{align}

\subsection{Unattributed edge case}. 

Let $s_i, y_i, z_i, \gamma_0, \gamma_1$ and $\gamma_2$ be defined as above and 
\begin{align}
N_0 = \sum_{i=1}^N (1-s_i) + s_i(1-z_i)(1-y_i),\\
N_1 = \sum_{i=1}^N s_iy_i(1-z_i) +s_iz_i(1-y_i),\\
N_2 = \sum_{i=1}^N s_iy_i z_i.
\end{align}

The E-step requires the expected likelihood after integrating out the latent positions. 
Let $\eta_0$ and $\eta_1$ be the first $K$ elements of $\alpha_0$ and $\alpha_1$ respectively:
\begin{align}
\eta_0 = (\alpha_{0, 1}, \ldots, \alpha_{0,K})'\\
\eta_1= (\alpha_{1, 1}, \ldots, \alpha_{1,K})'.
\end{align}
The E step estimates for $\tau_1, \tau_2$ and $\mathbf{v}$ can then be found using
\begin{align}
p(\mathbf{e}  |\alpha_1, \alpha_2, \tau_1, \tau_2, \mathbf{v}) \propto \prod_{i=1}^N \frac{\eta_{s_iy_i}\cdot \eta_{s_iy_i}}{\overline{\alpha}_{s_iy_i} \overline{\alpha}_{s_iz_i}}
\Big{(}1-\frac{\eta_0\cdot \eta_0}{\overline{\alpha}_0^2}\Big{)}^{(\gamma_0-N_0)} 
\Big{(}1-\frac{\eta_0\cdot \eta_1}{\overline{\alpha}_0 \overline{\alpha}_{1}}\Big{)}^{(\gamma_1-N_1)} 
\Big{(}1-\frac{\eta_1\cdot \eta_1}{\overline{\alpha}_1^2}\Big{)}^{(\gamma_2-N_2) }. \label{likelihood_unat}.
\end{align}

Let $A $ be the $n \times n$ multiadjacenty matrix associated with  $\mathbf{e}$. $A$ can be decomposed into matrices $\check{A}$ and $\dot{A}$ such that $A=\check{A} +\dot{A}$ where
\begin{align}
\check{A}_{u,v} = \sum_{i=1}^N \mathbf{1}_{u_i=u, v_i = v}((1-s_i) + s_i(1-y_i)(1-z_i))\\
\dot{A}_{uv} = A_{uv} - \check{A}_{uv}.
\end{align} 
and

The CM-step estimate  $\hat{\alpha}^{j+1}$ is based on estimates of $X$, the $K \times n$ matrix of latent positions, computed via an iterated algorithm based on gradient ascent. The starting point of the algorithm  is the least squares estimate of $X$ from the SVD procedure described in Section \ref{sec:fitting}. From this, the MLEs of $\alpha_0$ and $\alpha_1$ are computed. At each iteration of the algorithm, the new estimates of $X_i, i=1 \ldots n$ are then updated sequentially in the direction of the local maxima along the gradient of $g(X) = p(X|e, \alpha_0,\alpha_1, \tau_1, \tau_2, \mathbf{v})$:\\

for i $ \in \mathbf{v}^{j+1}$
\begin{gather*}
\Delta(g(X_i)) = (\dot{P}_i^T *X^T+ (\eta_1^T-1)/X^T_i)/N ,
 \end{gather*}
 
 for i $\in \{1, \ldots, n\} / \mathbf{v}^{j+1}$
\begin{eqnarray}
\Delta(g(X_i)) = (\check{P}_i^T *X^T+(\eta_0^T-1)/X^T_i)/N, 
\end{eqnarray}
where $\check{P}_i$ and $\dot{P}_i$ are  the $i$th columns of the matrices  $\check{P} = \check{A}_i/ (X^TX)$ and $\dot{P} = \dot{A}_i/ (X^TX)$, respectively, and `` / " denotes elementwise division. After each update of $X$, the MLEs of $\alpha_0$ and $\alpha_1$ are recomputed using the relevant latent positions, i.e. those corresponding to  draws from $f(\alpha_0)$ and $f(\alpha_1)$, based on the current estimates of $\tau_1, \tau_2$ and $\mathbf{v}$.

\bibliographystyle{asa}   
\bibliography{lucybib}  

 \end{document}